# Open nonradiative cavities as millimeter wave single-mode resonators

G. Annino[@], M. Cassettari, M. Martinelli
*Istituto per i Processi Chimico-Fisici del CNR, via G. Moruzzi 1, 56124 Pisa (Italy)*
(Dated: October 3, 2005)

**Abstract**

Open single-mode metallic cavities operating in nonradiative configurations are proposed and demonstrated. Starting from well-known dielectric resonators, possible nonradiative cavities have been established; their behavior on the fundamental $TE_{011}$ mode has been predicted on the basis of general considerations. As a result, very efficient confinement properties are expected for a wide variety of open structures having rotational invariance. Test cavities realized having in mind practical millimeter wave constraints have been characterized at microwave frequencies. The field distribution of some relevant configurations has been modeled by means of a finite-element numerical method. The obtained results confirm the expected high performances on widely open configurations. A possible excitation of the proposed resonators exploiting their nonradiative character is discussed, and the resulting overall ease of realization enlightened in view of millimeter wave employments.

---
[@] e-mail address: geannino@ipcf.cnr.it

# 1. Introduction

The millimeter wave frequency interval still represents a relatively unexplored field of investigation for electromagnetic spectroscopy. This was mainly due to the lack of proper sources having acceptable combinations of power, stability and tunability. A significant limitation to the development of millimeter wave spectroscopy seemed also due to the inadequateness of the waveguide-based propagation, solved in favor of sophisticated quasi-optical techniques. Recently, the solid-state technology developed first versatile millimeter wave sources, although still relatively limited in the emitted power. As a consequence, the microwave electric or magnetic field available on the sample is often too limited, requiring a resonant structure in order to increase the density of the electromagnetic energy in the active region of the spectrometer.

At millimeter wavelengths the most common resonant devices are given by overmoded resonators, which are relatively easy to be realized. In addition to the well-known Fabry-Perot cavities [1-4], more recently overmoded dielectric resonators have been proposed as well [5 and reference therein]. The quality factor $Q$ achievable in the above overmoded devices can be quite high (tens of thousand or higher at room temperature), leading to extremely selective frequency behaviors. On the other hand, in many spectroscopic applications the absolute sensitivity is related to the intensity of the electric or magnetic field on the sample; these quantities are in turn proportional to the ratio between the merit factor $Q$ and the active volume of the resonator. In this respect overmoded structures do not show clear advantages with respect to properly realized single-mode resonators and often, as in case of Fabry-Perot having reasonable dimensions, the achievable absolute sensitivity is clearly unsatisfactory. The use of overmoded resonators can be moreover hindered by geometrical constraints imposed by the employed setup. The above reasons basically explain the efforts made towards the realization of millimeter wave single-mode resonators [6, 7]. The main difficulty to be overcome in this case is given by the typical size of the employed components that, being of the order or smaller of the resonant wavelength, becomes harder and harder to be handled. In case of the popular $TE_{011}$ cylindrical cavity, for instance, the axial cross section that ensures the highest $Q$ factor is given by 3.95 mm by 3.95 mm at 100 GHz [8]. A much more critical point is represented by the excitation mechanism. The typically adopted solution, the coupling hole, is directly borrowed from the microwave technology. However, at 100 GHz a typical diameter of the coupling hole is given by 0.8 mm, with a corresponding thickness of the cavity wall around this aperture less than 0.05 mm; as a consequence, a very sophisticated design and realization of the cavity results [6]. Another key point of a single-mode cavity is given by its capability to allow a non-destructive access to the sample, namely an access that does not reduce sensibly the performances of the cavity itself. Such access can be necessary for the manipulation of the sample or its additional irradiation.

An exemplary case where these aspects are of basic importance is given by the Electron Paramagnetic Resonance (EPR) spectroscopy [9]. Here, the full rotation of the sample with respect to the magnetic field is generally required in single crystal studies. In addition, the cavity should be to some extent open, in order to allow an efficient static magnetic field modulation as well as radiofrequency excitation (for electron-nuclear double resonance techniques [9]), or optical access (in case of optically-activated samples or optically-detected magnetic resonance spectroscopy [10]). In the limit of high-field (millimeter wave) EPR, these requirements impose strict constraints on the realization of the cavity. As a result, the performances of the cavity or the capabilities of the employed spectroscopic techniques are often sensibly reduced in comparison to the ideal ones. In case of a $TE_{011}$ cylindrical cavity working at 275 GHz, the radiofrequency irradiation of the sample has been recently obtained by cutting 3 slits of 0.1 mm width perpendicularly to the axis of the cavity [11], following the solution adopted at 100 GHz [6]. The realization of single-mode resonators for millimeter wave spectroscopy seemed then mainly oriented to a simple rescaling of the design developed at microwave frequencies. Apparently, only limited efforts were dedicated to the realization of innovative solutions suitable for millimeter wave applications in any of their aspects. A recent attempt to realize single-mode dielectric resonators specifically conceived for millimeter wavelengths is reported in Ref. [12]. Here the state of the art of the power-to-field conversion efficiency at room temperature has been established, using low-permittivity resonators in the so-called NonRadiative (NR) configuration [13]. The peculiarity of these devices is given by their partially open structure, in which the metallic shielding is reduced to two planar and parallel mirrors; the distance between the mirrors can be as high as $\frac{\lambda_0}{2}$, where $\lambda_0$ is the wavelength in vacuum corresponding to the resonance frequency. The excitation of the NR Dielectric Resonators (NRDRs) can be obtained exploiting their nonradiative character, which basically prevents the propagation of the radiation outside the dielectric region [12]. A simple excitation configuration is then compatible with the structure of the resonator, without need of sophisticated *ad hoc* designs. The resulting device, based essentially on a dielectric disc and planar mirrors, is at the same time effective and easy to be realized, a mandatory characteristic for generalized millimeter-wave employments.

The main aim of this work is the generalization of the above NR configuration, in order to include purely metallic single-mode cavities in a simple and unified approach. The starting point will be represented by a possible analogy among NR dielectric configurations and metallic cavities. Following this analogy, the $TE_{011}$ mode of an Open NR Cavity (ONRC) will be investigated at microwave frequencies for a wide range of geometrical configurations. The obtained results will demonstrate the possibility of simple and innovative millimeter wave ONRCs.

This paper is structured as follows. Section 2 will introduce the theoretical background related to NR devices. Some generalized NR dielectric configurations will be introduced, and the analogous metallic cavities proposed. The expected behavior of their $TE_{011}$ mode will be discussed on the basis of general considerations. Section 3 will be dedicated to the experimental characterization. The criterion employed for the design of the test cavity resonating at microwave frequencies will be discussed. The results obtained on the $TE_{011}$ mode will be presented and analyzed. The implications of these results will be the topic of Section 4. Here the achieved merit factors will be interpreted in terms of the different loss mechanisms, and their extension to higher frequencies discussed. The field distributions of two relevant configurations, obtained with the aid of a finite-element numerical modeling, will be shown and discussed. Specific applications of these devices and their general perspectives will be discussed as well.

## 2. Generalities

The simplest configuration of a NRDR is given by a cylindrical low-loss dielectric rod having circular cross section, whose planar surfaces are in contact with conducting planes, as sketched in Fig. 1a. Here *l* represents the distance between the conducting planes, and *d* the diameter of the dielectric disc; the real part of the permittivity of the dielectric region will be indicated as $\varepsilon'$. A proper design of this structure ensures resonance modes whose merit factor is not affected by irradiation losses, that is nonradiative resonances. The NR regime is given by the condition $l < \frac{\lambda_0}{2}$, where $\lambda_0$ refers to each specific resonance frequency [12, 14].

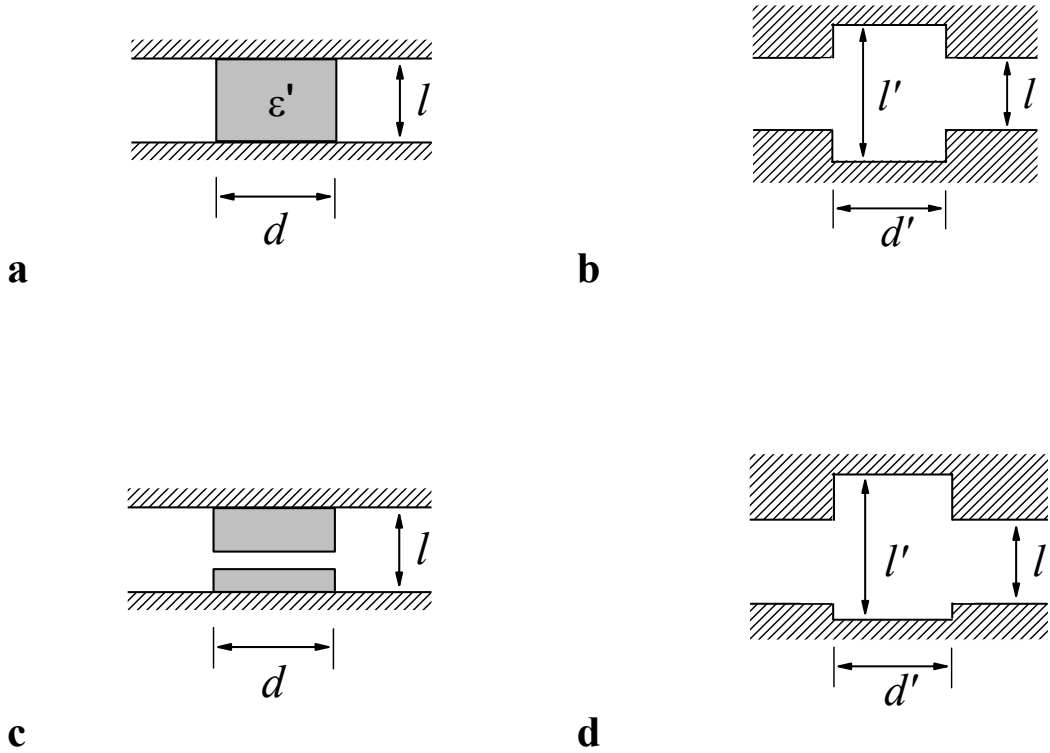

FIG. 1. *Axial cross section view of some NRDRs together with their metallic cavity analog.* ***a*** *Basic NRDR.* ***b*** *ONRC corresponding to the basic NRDR.* ***c*** *Split NRDR.* ***d*** *ONRC corresponding to the split NRDR.*

The working principle of the NRDRs can be naively explained in terms of allowed (above cutoff) and forbidden (below cutoff) propagation regimes. Indeed, when the conditions $l < \frac{\lambda_0}{2} < l \cdot \sqrt{\varepsilon'}$ and $d \sim l$ are fulfilled the radiation can propagate inside the dielectric region (which is above cutoff), and cannot propagate in the outer region (below cutoff), where the fields decay exponentially. The quantity $l \cdot \sqrt{\varepsilon'}$ represents the optical thickness of the dielectric region. In a slightly different formulation, the NRDR operates below the cutoff frequency of the parallel-plates TE and TM modes, while the coupling to the cutoff-less TEM modes vanishes ([12] and reference therein). The fundamental modes of low-permittivity NRDRs have been investigated up to millimeter wavelengths [12, 13].

The NR configuration of Fig. 1a can be generalized as shown in Fig. 1c ('split' NRDR) and Fig. 2a ('tuned' NRDR). Both these variants have been characterized at microwave frequencies, as anticipated in Ref. [13]; high $Q$ values were obtained on the $TE_{011}$ and $TM_{011}$ modes, as well as on hybrid modes. These results will be presented in detail elsewhere. Similar structures have been investigated in an interesting paper by Kobayashi and coworkers [15]. Another possible generalization of the basic NR configuration is shown in Fig. 2c ('composite' NRDR).

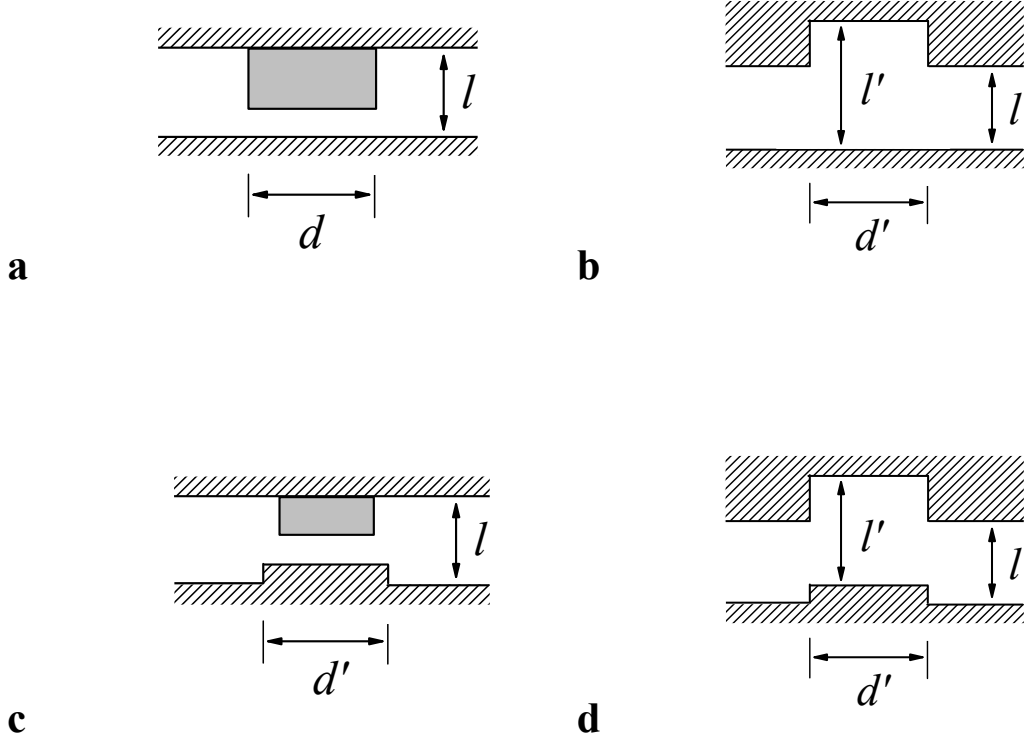

FIG. 2. *Axial cross section view of some NRDRs together with their metallic cavity analog.* **a** *Tuned NRDR.* **b** *ONRC corresponding to the tuned NRDR.* **c** *Composite NRDR.* **d** *ONRC corresponding to the composite NRDR.*

The point of interest is here given by the determination of possible analogies among these NRDRs and purely metallic configurations. Following the above-discussed working principle, the necessary condition for a NR device is the realization of a region of allowed propagation surrounded by a region of forbidden propagation. A NR dielectric configuration can be formally transformed in a similar NR metallic structure by replacing the dielectric regions with cavities having similar optical size. The metallic configurations shown in Figs. 1b, 1d, 2b and 2d can be then considered to some extent analogous to the NRDRs of Figs. 1a, 1c, 2a and 2c, respectively; similar electromagnetic behaviors are in turn expected. The working condition $l < \frac{\lambda_0}{2} < l \cdot \sqrt{\varepsilon'}$ is now replaced by $l < \frac{\lambda_0}{2} < l'$, where as usual the height $l'$ of the metallic cavity is assumed of the order of its diameter $d'$. Following this simplified model, open single-mode metallic resonators should be possible also for slight local enlargements of the distance between two parallel conducting plates. In analogy to the behavior observed in NRDRs, for distances $l$ sensibly lower than $\frac{\lambda_0}{2}$ the NR condition is well verified and the evanescent field outside the central region decreases very rapidly. On the contrary, when $l$ approaches $\frac{\lambda_0}{2}$ the confinement mechanism becomes weaker and weaker and the resonance mode extends well beyond the central region, requiring large conducting plates to prevent irradiation losses.

For some modes of high practical importance the above analysis can be formulated on the basis of more general and rigorous considerations. It is indeed well known that in rotationally invariant dielectric structures with isotropic or uniaxial complex permittivity the modes having azimuthal invariance (vanishing azimuthal index) are either transverse electric TE or transverse magnetic TM. These modes can be accordingly indicated as $TE_0$ and $TM_0$. The demonstration of this property is outlined in Appendix. The $TE_0$ modes of a NR structure do not share in particular any field components with the cutoff-less TEM modes, so that their irradiation losses in principle vanish for any configuration fulfilling the nonradiative criterion. Accordingly, any NR structure having rotational invariance shows at least a family of modes characterized by vanishing irradiation losses. This is equally true for the particular case of purely metallic structures. The $TE_0$ modes of the open configurations reported in Figs. 1b, 1d, 2b and 2d are then expected free from irradiation losses until their resonance frequency violates the nonradiative criterion, provided that the extension of the metallic plates is large enough. The remaining part of the paper will be basically dedicated to the confirmation of this conclusion.

The use of the $TE_{011}$ mode of the symmetrically open configuration of Fig. 1b has been proposed for the dielectric characterization of planar materials [16-18]. The mathematical modeling of this device has been elegantly developed in the paper by Janezic and Baker-Jarvis [19]. However, the analysis and the use of this configuration were limited to small gaps, in which the electromagnetic fields extend only marginally in the volume outside the central cavity. The limit height of the gap, its allowed positions along the body of the cavity, and the possible benefits of widely open structures were apparently not recognized.

The main practical aspect to be faced with for an effective use of the resonator concerns its excitation. In principle, any of the well-established excitation techniques as coupling holes [20], loops or probes [9], dielectric waveguides in NR configurations [21], can be employed in combination with the proposed ONRCs. However at millimeter wavelengths, where these devices should express their full benefits, the only practicable solution seems given by coupling holes. As anticipated, the use of coupling holes requires a very critical and sophisticated design of the resonator, to which often correspond limited adjustments of the coupling parameter. In our case a much more versatile solution exploits the NR character of the proposed devices, in which the electromagnetic fields outside the central region decrease exponentially. These evanescent fields can be overlapped to those of the incoming radiation by limiting in a proper way the extension of the metallic walls, as shown in Ref. [12]. The coupling parameter can be adjusted by changing the relative position between the resonator and the incoming radiation. This solution, which has been proven quite simple and effective also at millimeter wavelengths [13], can be directly adopted for ONRCs in virtue of their analogy with NRDRs. In case of ONRCs, an alternative control of the coupling parameter can be obtained by changing the distance $l$ between the conducting plates; in this way the intensity of the evanescent field can be effectively controlled. The resulting perturbation of the resonance frequency can be compensated acting on the height of the central cavity.

### 3. Experimental evidences

The cavity employed for the microwave measurements is sketched in Fig. 3. It was basically composed by two aluminum plates, each of them equipped with a mobile aluminum plunger, separated by three pedestals. The measurements were intended for the demonstration of the NR principle as well as for a realistic evaluation of the achievable millimeter wave performances; accordingly, higher mechanical tolerances than that easily obtainable were intentionally chosen. The upper part of the cavity was given by a parallelepiped having a 30 mm by 30 mm cross section and a 13.85 mm diameter axial hole. The lower plate was realized in a similar way, with a cross section of 40 mm by 40 mm, slightly larger than the upper one for mechanical convenience; the diameter of its hole was 14.35 mm.

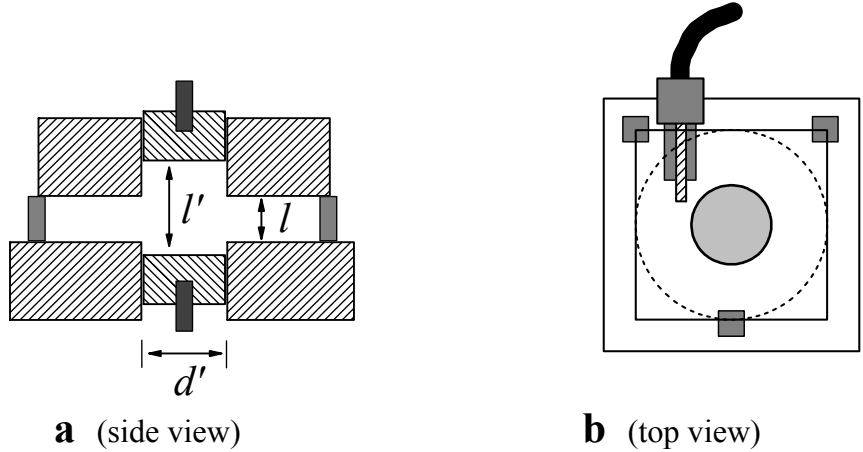

FIG. 3. *Metallic cavity employed in the measurements.* ***a*** *Axial cross section view.* ***b*** *Top view. The dashed circumference indicates the limit extension of the active region.*

The plungers were given by discs with 13.7 mm diameter and 10 mm height. The distance between the plates was fixed to $4.25 \pm 0.1$ mm. The position of the plungers was changed independently from each other; in this way the aperture of the cavity was freely positioned along its axis. None of the metallic surfaces were polished. The resonances were detected in a reflection configuration by means of a vector network analyzer. The excitation was guaranteed by a probe connected to the analyzer through a flexible cable; for the purpose of the present analysis this indeed represents the most versatile solution at microwave frequencies.

Following the analysis of the previous section, the maximum working frequency of the employed cavity is given by $35.3 \pm 0.8$ GHz, as obtained from the limit condition $l = \frac{\lambda_0}{2}$. Close to this limit most of the electromagnetic energy is distributed on a large volume outside the central cavity. The related energy density is then correspondingly low; in addition, a relevant contribution to the losses is now expected from the finite conductivity of the mirrors. As previously discussed, this often means a low gain in the absolute sensitivity. The structure of the test cavity chosen for these measurements allowed the monitoring of the maximum spread of the resonance mode. The NR confinement was indeed guaranteed as far as the rim of the upper plate, namely for a maximum radius about twice to that of the central cavity, as indicated in Fig. 3b. Accordingly, when the active volume of the resonance became about three times that of the central cavity, a sensible reduction of the measured merit factor was expected, as marker of the low energy density limit.

The measurements were organized in three different series, in which one plunger was fixed in a definite position and the other one moved in order to tune the cavity. In this way the same resonance condition was obtained for three different positions of the aperture along the axis of the resonator. The $TE_{011}$ mode was unambiguously identified among the different observed resonances by comparing its frequency with that calculated in the corresponding close cavity [22]. Indeed, this mode is expected only slightly perturbed by the aperture when the NR condition is well verified and the height of the cavity $l'$ is sensibly higher than $l$.

In the first series of measurements the upper plunger was fixed 7 mm above the aperture (this configuration is hereafter referred to as '7 mm configuration'). The obtained resonance frequency and unloaded merit factor $Q_0$ are reported in Fig. 4 as a function of the position of the lower plunger with respect to the aperture; positive values refer to positions inside the aperture, as shown in Fig. 2d. The maximum protrusion of the plunger was limited by the presence of the probe.

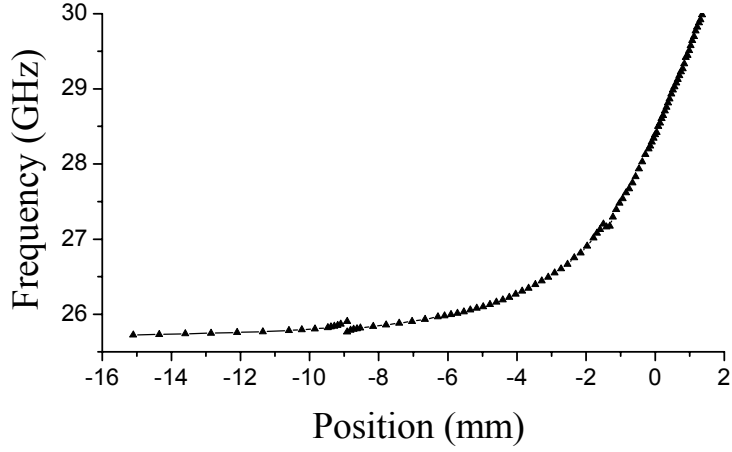

a

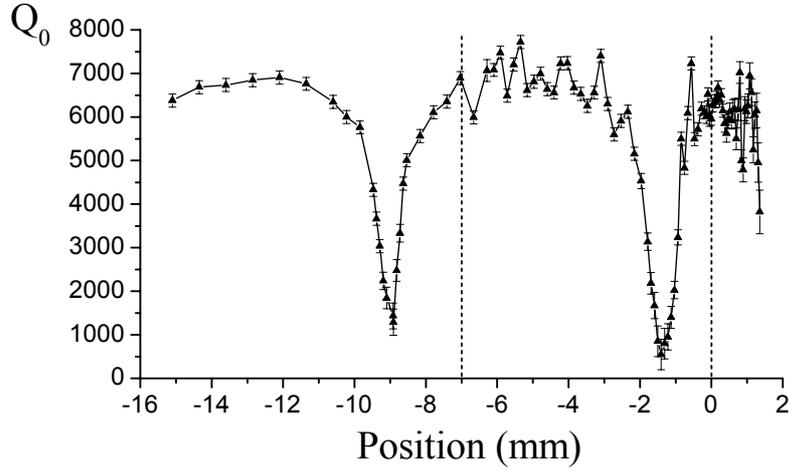

b

FIG. 4. *Measurements on the $TE_{011}$ mode of the 7 mm configuration (triangles), as described in the text.* ***a*** *Resonance frequency vs. the position of the lower plunger.* ***b*** *Unloaded merit factor vs. the position of the lower plunger. The dashed lines indicate the conditions of symmetric cavity and cavity with a planar mirror. The error bars are reported.*

The second and third series were obtained by fixing the upper plunger 5 mm and 3 mm above the aperture (hereafter referred to as '5 mm configuration' and '3 mm configuration'); the related data are reported in Fig. 5 and Fig. 6, respectively. In Figs. 4b, 5b and 6b the vertical lines indicates the conditions of symmetric cavity and cavity with one planar mirror, as shown in Fig. 1b and 2b, respectively. Fig. 7 reports the merit factors of all measurements versus the total height $l'$.

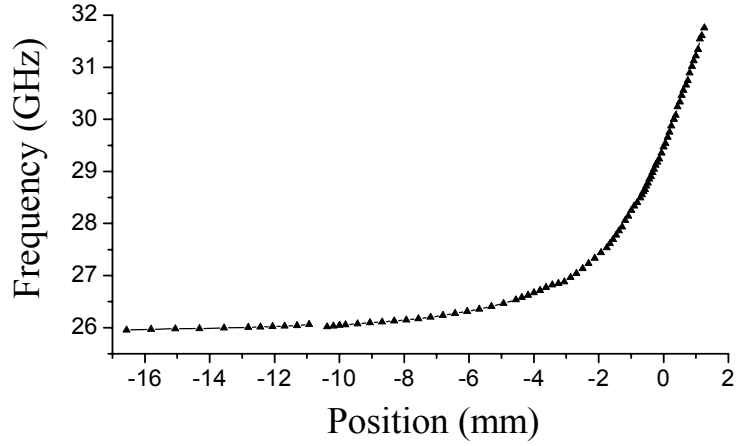

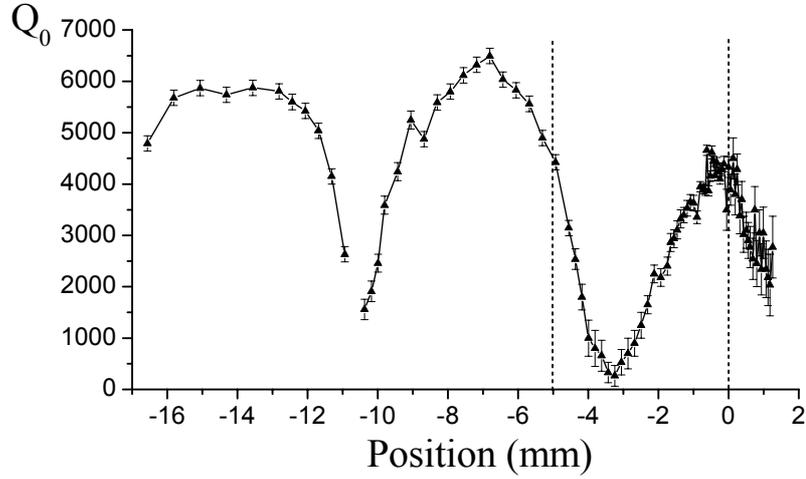

FIG. 5. *Measurements on the $TE_{011}$ mode of the 5 mm configuration (triangles), as described in the text.* ***a*** *Resonance frequency vs. the position of the lower plunger.* ***b*** *Unloaded merit factor vs. the position of the lower plunger. The dashed lines indicate the conditions of symmetric cavity and cavity with a planar mirror. The error bars are reported.*

In all series of measurements a similar behavior is observed. The tuning of the $TE_{011}$ mode increases exponentially as the height of the cavity is reduced, as expected. The measured merit factor is generally quite high, excluding two dips that appear in any series. The origin of these dips can be inferred from the obtained data. The first dip is related to an anticrossing phenomenon with a second mode, as evidenced by the typical frequencies repulsion behavior. The spurious mode is not visible outside the interaction region and is very likely characterized by a low merit factor. In the terminology of the lumped circuit representation this phenomenon is typically referred to as reactive coupling of modes [23-26]. The second dip does not show a clear anticrossing behavior, so it can be either related to a crossing of modes (resistive coupling of modes), or to a leakage phenomenon through the aperture.

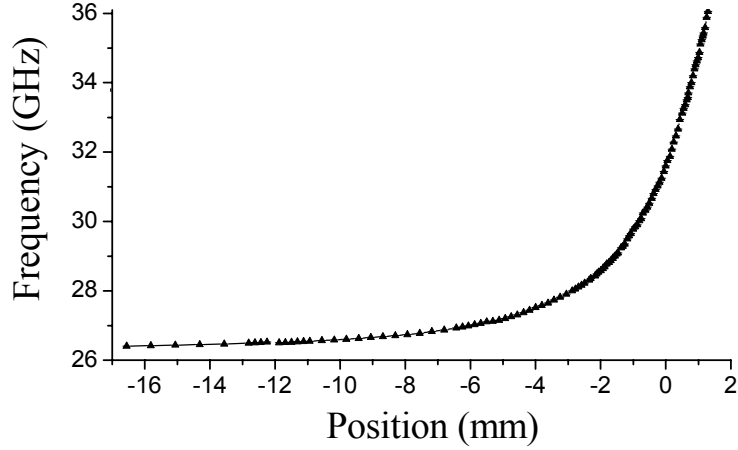

a

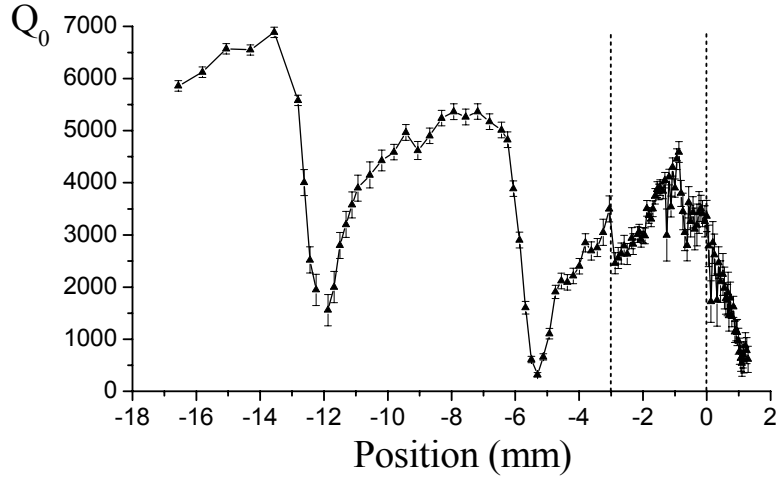

b

FIG. 6. *Measurements on the $TE_{011}$ mode of the 3 mm configuration (triangles), as described in the text. **a** Resonance frequency vs. the position of the lower plunger. **b** Unloaded merit factor vs. the position of the lower plunger. The dashed lines indicate the conditions of symmetric cavity and cavity with a planar mirror. The error bars are reported.*

In the latter case however the irradiation losses of the three series of measurements should differ remarkably; indeed, the diverse position of the aperture along the cavity in correspondence to the dip should give sensibly different projections of the resonance mode on the leaking modes. A similar argument suggests different positions for the minimum of the dip when plotted in terms of the total height $l'$ of the cavity. On the contrary, the close coincidence of the position of the dip shown in Fig. 7, as well as its almost constant height resulting from Figs. 4b, 5b and 6b, appear as decisive evidence in favor of a modal interaction effect. Also in this case the spurious mode is not directly visible and is expected to be characterized by a low merit factor.

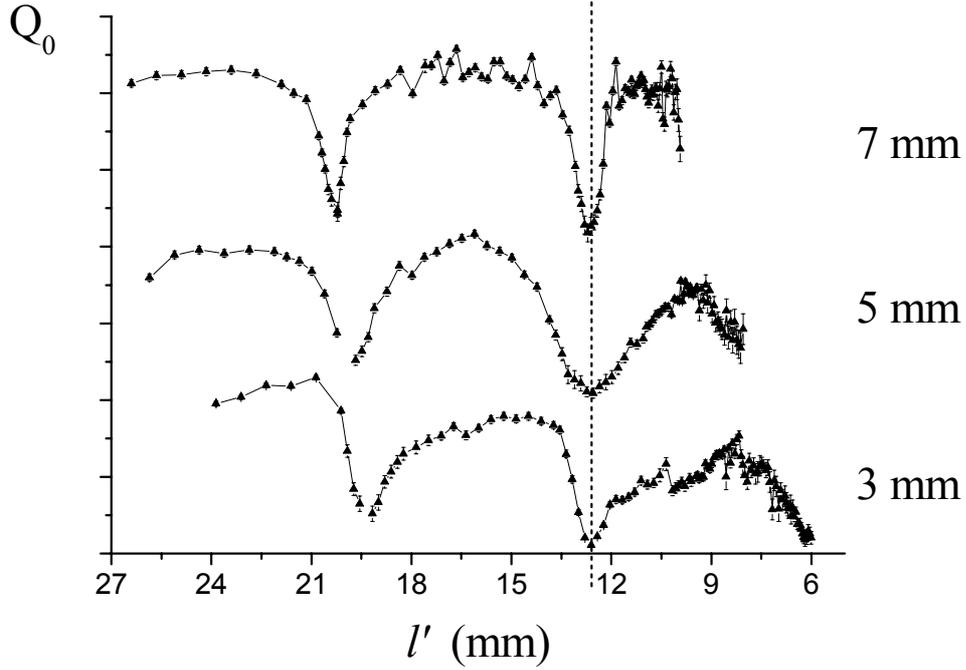

FIG. 7. *Unloaded merit factors obtained for the configurations described in the text vs. the total height l' of the cavity. The curves related to the 7 mm and 5 mm configurations have been translated vertically. The error bars are reported.*

The 3 mm configuration allowed the analysis of the whole frequency interval predicted by the NR condition. The maximum observed resonance frequency was about 36 GHz, in full agreement with the above analysis. Accordingly, the expected nonradiative behavior seems fully confirmed in a wide range of conditions.

### 4. Discussions and conclusions

The obtained data suggest further considerations. Combining the results of all the investigated configurations, it follows that the NR confinement is effective for any of the geometries shown in Figs. 1b, 1d, 2b, 2d. In particular, in Fig. 6b the part of the curve comprised between the second dip and the NR limit crosses all the above geometries. The importance of this region is clearly evidenced by Fig. 8, where the $Q_0$ factor of the 3 mm configuration is plotted versus the related resonance frequency. This figure shows in particular that the largest part of the tuning frequency range is free from serious deteriorations of the merit factor. The highest $Q_0$ beyond the second dip, $Q_{0,m} \sim 4500$, is obtained around 30 GHz for a total height $l' \sim 0.6 \cdot d' \sim 8.3\text{mm}$. In this condition the aperture of the cavity represents about 50% of its total height. If the limit of acceptable $Q_0$ values is fixed to 3000, the corresponding interval of acceptable frequencies extends for about 10% around 30 GHz. Better merit factors and wider tuning intervals are expected for optimized cavity designs. For instance, the losses in the metallic walls can be very likely reduced assuming for a given frequency a cross section close to the optimal one $l' = d'$ calculated for the close cavity [8].

The level of the irradiation losses can be estimated from the measured $Q_0$ by using the well-known expression $\frac{1}{Q_0} = \frac{1}{Q_{0,met}} + \frac{1}{Q_{0,rad}}$ [27]. Here $Q_{0,met}$ and $Q_{0,rad}$ are the merit factors calculated taking into account only metallic and irradiation losses, respectively. The typical $Q_0$ obtained when the aperture of the cavity is small in comparison to its total height is of the order of 6000. By assuming that in these conditions the irradiation losses are negligible and that the losses in the metallic walls do not change appreciably by tuning the cavity, $Q_{0,rad}$ can be estimated for configurations having large apertures.

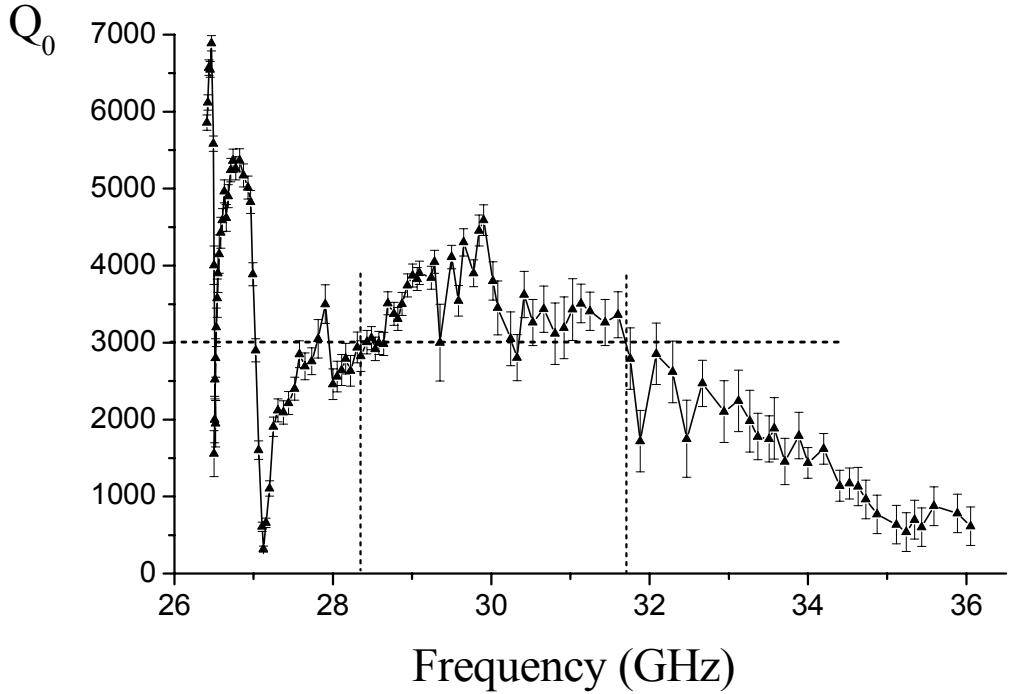

FIG. 8. *Unloaded merit factor of the 3 mm configuration vs. the related resonance frequency. The error bars are reported.*

In the above frequency interval around 30 GHz the value of $Q_{0,rad}$ estimated in this way is higher than 6000. This evaluation is expected meaningful also at millimeter and submillimeter wavelengths, since the irradiation losses are in principle independent from the resonance frequency when the size of the cavity is just rescaled [12]; in addition, the relative mechanical tolerances of the employed cavity are easily achievable at frequencies sensibly higher than the microwave ones. The behavior of the irradiation losses confirms in general that the electromagnetic fields were quite weak on the boundary of the resonator for most of the investigated geometries, according to the discussion on the role of the limited extension of the metallic plates. The resonance mode is generally well confined close to the central cavity, ensuring a limited active volume and then a relevant energy density.

These qualitative results were confirmed by a numerical analysis of the $TE_{011}$ field distribution of the employed ONRCs. Two representative configurations were in particular investigated by means of the finite-element software FEMLAB 3.0a (COMSOL, Sweden): the symmetric cavity shown in Fig. 1b, and the cavity with one planar mirror shown in fig. 2b, both in the case of the 3 mm configuration. The azimuthal electric field on the axial cross section of the symmetric cavity is reported in Fig. 9. Only half of the section of the cavity is reported, due to the rotational invariance on the mode. Although the lateral aperture represents in this case about 40% of the total height of the cavity, only about 5% of the electric energy and about 12% of the magnetic energy is stored outside the central volume, namely in the lateral aperture. Fig. 10 shows the azimuthal electric field in the case of cavity with one planar mirror, where the lateral aperture represents about 60% of the total height.

The energy stored outside the central volume is now given by an electric component representing about 8% of the total electric energy, and by about 13% of the magnetic energy.

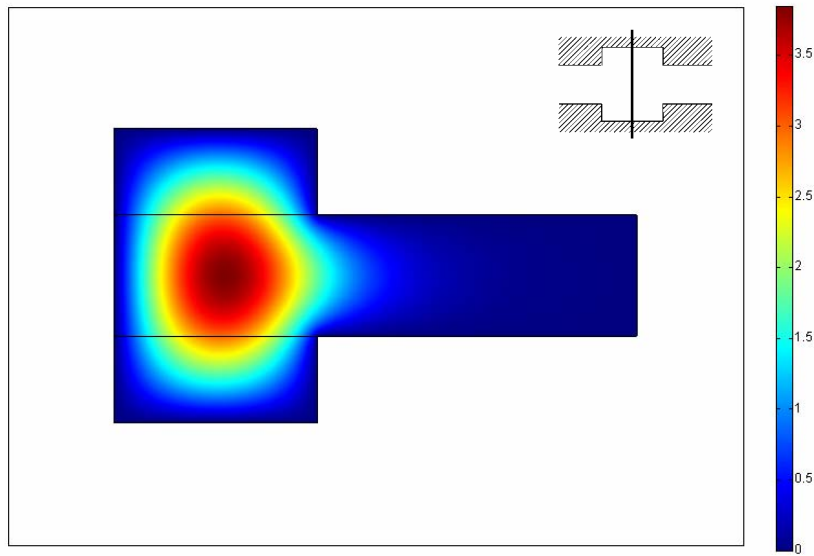

FIG. 9. *Azimuthal electric field distribution on the axial cross section of the 3 mm configuration, in correspondence of the symmetric geometry indicated in Fig. 6. Only half of the section is reported, due to the rotational invariance on the mode, as schematically shown in the inset. On the right the colors scale, expressed in arbitrary units.*

In both cases, in addition, the electromagnetic energy outside the central volume decay quite rapidly with the distance. The above conditions mark the limit of the allowed working region indicated in Fig. 8. All the geometric configurations belonging to this region are therefore characterized by an almost complete confinement of the radiation in the central volume; the conversion factor is then expected comparable to that of the corresponding close cavity.

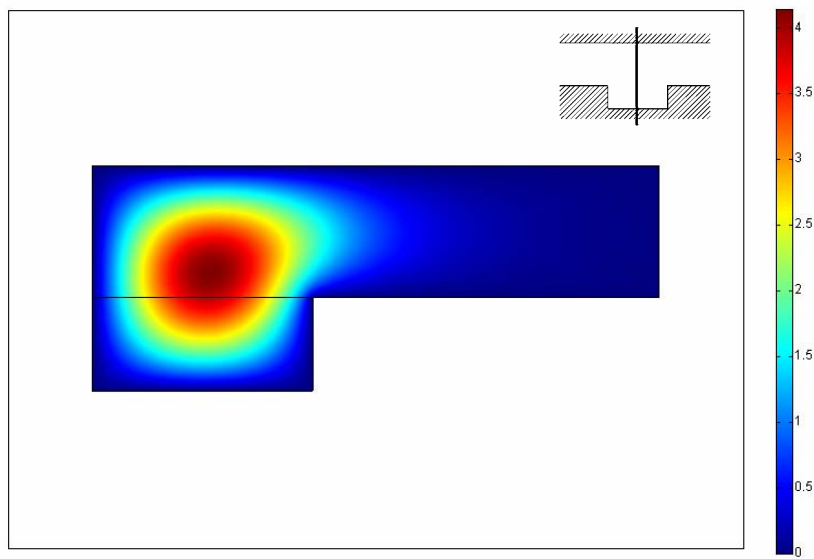

FIG. 10. *Azimuthal electric field distribution on the axial cross section of the 3 mm configuration, in correspondence of the geometry with a planar mirror indicated in Fig. 6. Only half of the section is reported, due to the rotational invariance on the mode, as schematically shown in the inset. On the right the colors scale, expressed in arbitrary units.*

The above analysis demonstrates that ONRCs working on the $TE_{011}$ mode can be easily realized. In particular, the aperture between the two parts of the resonator can be largely varied in width and position without severe degradations of its performances, allowing the realization of widely open single-mode resonators having moderately extended metallic plates. The same conclusion is expected true for any $TE_0$ mode. The accurate knowledge of the whole field distribution allows in particular further spectroscopic applications, as the complex dielectrometry of laminar materials.

The proposed analogy between NRDRs and ONRCs suggests a possible merging between these two classes of resonant structures, in which dielectric regions are combined with open cavities following the indications of the NR confinement mechanism. A rich variety of open single-mode resonators can be realized accordingly; the resulting configurations can share the typical benefits of their constituents, as the high energy density of NRDRs and the ease of realization and tuning of ONRCs. The mechanical access to the cavity can be ensured for instance by a low-loss dielectric plate, inserted in the cavity through its planar aperture. The dielectric plate can extend well beyond the central resonant region, provided that the NR condition is verified throughout. In this way a high degree of freedom can be achieved in the manipulation of the sample.

In conclusion, the expected behavior of the $TE_{011}$ mode of ONRCs is fully confirmed. The excitation of these resonators can be easily obtained also at millimeter wavelengths by exploiting their NR character. The overall structure of the proposed devices appears then inherently easy to be realized and operated, and shows clear advantages with respect to standard single-mode cavities. The performances suggested by the present analysis are quite promising, and typical of a wide variety of rotationally invariant metallic, dielectric or composite configurations obeying only to the NR condition. The better expression of these benefits is expected at millimeter and submillimeter wavelengths, where open NR single-mode resonators can represent a decisive innovation in many applications. The potentialities of these resonators seem still largely unexplored, requiring a further theoretical and experimental work in view of their full realization.

**Appendix**

In a linear and local medium with rotational invariance about the z-direction, the most general expressions for the tensors of dielectric permittivity and magnetic permeability are given by

$$\begin{pmatrix} \varepsilon_\perp & 0 & 0 \\ 0 & \varepsilon_\perp & 0 \\ 0 & 0 & \varepsilon_z \end{pmatrix}, \qquad \begin{pmatrix} \mu_\perp & 0 & 0 \\ 0 & \mu_\perp & 0 \\ 0 & 0 & \mu_z \end{pmatrix}.$$

In any homogeneous region of this medium, the rotational Maxwell equations can be written in terms of two independent components. Choosing the axial fields as independent fields, the rotational equations can be expressed in cylindrical coordinates as [27, 28]

$$\left( \frac{\partial^2}{\partial z^2} + \frac{\omega^2}{c^2} \varepsilon_\perp \mu_\perp \right) H_\rho = j \frac{\omega \varepsilon_\perp}{c} \frac{1}{\rho} \frac{\partial E_z}{\partial \varphi} + \frac{\partial^2 H_z}{\partial z \partial \rho},$$

$$\left( \frac{\partial^2}{\partial z^2} + \frac{\omega^2}{c^2} \varepsilon_\perp \mu_\perp \right) E_\rho = \frac{\partial^2 E_z}{\partial z \partial \rho} - j \frac{\omega \mu_\perp}{c} \frac{1}{\rho} \frac{\partial H_z}{\partial \varphi},$$

$$\left( \frac{\partial^2}{\partial z^2} + \frac{\omega^2}{c^2} \varepsilon_\perp \mu_\perp \right) H_\varphi = -j \frac{\omega \varepsilon_\perp}{c} \frac{\partial E_z}{\partial \rho} + \frac{1}{\rho} \frac{\partial^2 H_z}{\partial z \partial \varphi},$$

$$\left( \frac{\partial^2}{\partial z^2} + \frac{\omega^2}{c^2} \varepsilon_\perp \mu_\perp \right) E_\varphi = \frac{1}{\rho} \frac{\partial^2 E_z}{\partial z \partial \varphi} + j \frac{\omega \mu_\perp}{c} \frac{\partial H_z}{\partial \rho}.$$

In the same manner, the solenoidal equations give

$$\varepsilon_\perp \frac{1}{\rho} \frac{\partial (\rho \cdot E_\rho)}{\partial \rho} + \varepsilon_\perp \frac{1}{\rho} \frac{\partial E_\varphi}{\partial \varphi} = -\varepsilon_z \frac{\partial E_z}{\partial z}$$

and

$$\mu_\perp \frac{1}{\rho} \frac{\partial (\rho \cdot H_\rho)}{\partial \rho} + \mu_\perp \frac{1}{\rho} \frac{\partial H_\varphi}{\partial \varphi} = -\mu_z \frac{\partial H_z}{\partial z}.$$

In case of solutions with azimuthal invariance ($\frac{\partial}{\partial \varphi} \equiv 0$), the above equations lead necessarily to TE modes with fields ($H_\rho$, $E_\varphi$, $H_z$) and to TM modes with fields ($E_\rho$, $H_\varphi$, $E_z$), which do not share any field component. These basic solutions, obtained for a specific homogeneous region, can be possibly mixed by the boundary conditions. A system with no source terms (free charges and currents), and media with finite conductivity will be assumed. In this manner, the description of the conductivity can be included in the model by means of a proper complex permittivity.

Let's consider now the boundary between two different homogeneous regions, labeled with 1 and 2, and the normal to the separating surface, labeled with $\hat{n}$. In the above assumptions the boundary conditions reduce to continuity conditions, of the form $\left\{ \vec{D}_2 - \vec{D}_1 \right\} \cdot \hat{n} = 0$ and $\left\{ \vec{B}_2 - \vec{B}_1 \right\}$

$$\left\{ \begin{array}{l} \vec{E}_2 - \vec{E}_1 \\ \vec{H}_2 - \vec{H}_1 \end{array} \right\} \times \hat{n} = \vec{0}.$$ Where defined, the vector $\hat{n}$ cannot have an azimuthal component, in virtue of the rotational invariance of the system. Accordingly, $\hat{n} = n_\rho \hat{\rho} + n_z \hat{z}$ at any point. The scalar boundary conditions lead therefore to the continuity of the term $\varepsilon_\perp E_\rho n_\rho + \varepsilon_z E_z n_z$ and of the term $\mu_\perp H_\rho n_\rho + \mu_z H_z n_z$, whereas the vector conditions lead to the continuity of the components $E_\varphi$ and $H_\varphi$, and of the terms $n_\rho E_z - n_z E_\rho$ and $n_\rho H_z - n_z H_\rho$. The boundary conditions can be separated again in a set of equations involving only the above TE field components and a set of equations involving only the above TM field components. As a consequence, these conditions do not mix the TE and TM modes obtained from the Maxwell equations, which represent therefore proper solutions for the whole configuration. The generalization of this result to systems with a continuous variation of permittivity and permeability can be obtained by approximating these quantities with stepwise functions and then taking a limit operation. Analogously, a limit operation on the conductivity of the employed materials allows the generalization of the above results to configurations including ideal conductors.

In conclusion, in a rotationally invariant configuration the $\varphi$ – independent electromagnetic solutions, both bounded and unbounded, are necessarily transverse electric or transverse magnetic, with nonvanishing components given by ($H_\rho$, $E_\varphi$, $H_z$) and ($E_\rho$, $H_\varphi$, $E_z$), respectively.

.